\documentclass[10pt,a4paper,article,oneside]{memoir}

\renewcommand{\OnehalfSpacing}{%
  \setSpacing{1.618}
}
\OnehalfSpacing
\setlrmarginsandblock{4.011cm}{4.011cm}{*}
\setulmarginsandblock{1.683cm}{1.683cm}{*}
\checkandfixthelayout

\counterwithout{section}{chapter}
\setsecnumdepth{subsection}

\usepackage[utf8]{inputenc}
\usepackage{libertinus-type1}
\usepackage{libertinust1math}

\usepackage{hyperref}

\usepackage{amsmath}
\usepackage{mathtools}
\usepackage{bm}
\usepackage{booktabs}
\usepackage{pgfplots}
\pgfplotsset{compat=1.18}

\DeclarePairedDelimiter{\parens}{(}{)}

\DeclarePairedDelimiter{\abs}{\lvert}{\rvert}
\DeclarePairedDelimiter{\norm}{\lVert}{\rVert}

\DeclareMathOperator{\perm}{perm}

\newcommand{\ket}[1]{\lvert #1 \rangle}
\newcommand{\nb}{\mathbf{n}}
\newcommand{\cb}{\mathbf{c}}
\newcommand{\eb}{\mathbf{e}}
\newcommand{\Ub}{\mathbf{U}}
\newcommand{\Jb}{\mathbf{J}}

\newcommand{\figwidth}{0.85\textwidth}
\newcommand{\figheight}{0.5\textwidth}
\newcommand{\figlegendfont}{\small}
\newcommand{\figmarksize}{2.5pt}

\title{The quantum multinomial distribution: a combinatorial formulation of multiphoton interference}
\author{%
  Alfonso Martinez\thanks{alfonso.martinez@upf.edu} \qquad
  Josep Font-Segura\thanks{josep.font-segura@upf.edu}\\[4pt]
  {\normalsize Department of Engineering, Universitat Pompeu Fabra, Barcelona, Spain}%
}
\date{}

\begin{document}

\maketitle

\begin{abstract}
This paper presents a quantum generalization of the multinomial distribution for the transition probabilities of $m$ identical photons in a $k$-port linear optical interferometer: two multinomial coefficients (one for the input configuration, one for the output) times the squared modulus of a coherent sum over routing matrices, weighted by the multivariate hypergeometric distribution; no Hilbert space formalism is needed to state or evaluate it. The classical multinomial is recovered when all photons enter through a single port, the coherent sum degenerating to a single term with no interference; the quantum family is not a generalization in the Askey sense but a parallel family that departs from classical statistics through the coherence of the amplitude summation. The $r$-th factorial moment carries a squared multinomial coefficient in place of the classical single one, the extra factor arising from the two copies of the amplitude expansion whose indices the Fock state forces to agree; for the beam splitter, the third cumulant is invariant under bosonic interference and the quantum departure first appears in the fourth cumulant as negative excess kurtosis; for multiport interferometers, however, three-body interference breaks this invariance and the departure enters already at the third cumulant. Cross-mode covariances involve the phases of the scattering matrix through coherence terms that strengthen output anti-correlations beyond the classical value; together with the squared-coefficient signature in the single-mode moments, these provide low-order statistical witnesses for boson sampling verification without requiring the full permanent computation.
\end{abstract}

\section{Introduction}\label{sec:intro}

The output photon-number statistics of a lossless linear optical network (a multiport interferometer described by a $k\times k$ unitary scattering matrix~$\Ub$; any such matrix can be realized by a network of beam splitters and phase shifters~\cite{reck1994}) are fundamental in quantum optics. Given $m$ identical photons distributed among $k$ input ports according to a Fock state~$\ket{\nb}$, with $\nb = (n_1, \ldots, n_k)$ a composition of~$m$, the probability $P(\cb \mid \nb)$ of detecting the output configuration $\cb = (c_1, \ldots, c_k)$ is given by the squared permanent of an $m \times m$ scattering submatrix~$\Ub_S$, divided by the product of the input and output occupation-number factorials~\cite{scheel2008}. The computational intractability of the permanent~\cite{valiant1979} underlies the boson sampling proposal~\cite{aaronson2013}. The exact evaluation of these probabilities, both for theoretical understanding and experimental prediction, has motivated a variety of equivalent formulations; see~\cite{tichy2014} for a comprehensive review.

For the two-port case ($k = 2$, the beam splitter), several equivalent expressions for $P(\cb \mid \nb)$ are available in the literature: Jacobi polynomials~\cite{campos1989}, hypergeometric functions~\cite{mandel1995}, and Wigner $d$-matrix elements from the angular momentum representation of SU(2)~\cite{yurke1986}. For the general multiport case, the permanent of the scattering submatrix remains the standard expression.

Each formulation has its merits. The Jacobi polynomial form is compact; the Wigner $d$-matrix form connects to the representation theory of SU(2); the permanent applies to arbitrary~$k$. However, none of these formulations makes the mechanism of quantum interference and its departure from classical particle statistics directly visible. The classical limit (all photons entering through a single port) is not recognizable as a special case; the role of photon indistinguishability in producing the interference is not isolated; and the combinatorial structure underlying the transition probabilities is hidden behind either special functions or a sum over $m!$~permutations.

In this paper, we present a reformulation that makes these features transparent. We express the transition probability as
\begin{equation}\label{eq:quantum-multinomial}
P(\cb \mid \nb) \;=\; \binom{m}{\nb}\binom{m}{\cb} \abs*{\sum_{\Jb \in \mathcal{J}(\nb,\cb)} w_{\Jb}\, a_{\Jb}}^2
\end{equation}
where $\binom{m}{\nb} = m!/\prod_{i=1}^k n_i!$ is the multinomial coefficient; the sum runs over the set $\mathcal{J}(\nb,\cb)$ of \emph{routing matrices} $\Jb = (J_{ij})_{i,j=1}^k$, that is, non-negative integer $k\times k$ matrices with row sums~$\nb$ and column sums~$\cb$, and the two remaining ingredients have transparent combinatorial meanings:
\begin{itemize}
\item The \emph{amplitudes} $a_{\Jb} = \prod_{i=1}^k \prod_{j=1}^k U_{ij}^{\,J_{ij}}$ are products of single-photon scattering amplitudes, one factor per routed photon.
\item The \emph{weights} $w_{\Jb} = \mu_{\Jb} / \binom{m}{\cb}$, with $\mu_{\Jb} = \prod_{i=1}^{k} \binom{n_i}{J_{i1}, \ldots, J_{ik}}$, form the multivariate hypergeometric distribution: the probability that $m$ labeled items belonging to $k$ categories of sizes $n_1, \ldots, n_k$ fall into $k$ bins of sizes $c_1, \ldots, c_k$ with contingency table~$\Jb$. The normalization $\sum_{\Jb \in \mathcal{J}} w_{\Jb} = 1$ is the multivariate Chu--Vandermonde identity.
\end{itemize}
The prefactor $\binom{m}{\nb}\binom{m}{\cb}$ counts the number of (input labeling, output labeling) pairs when $m$~identical photons are temporarily assigned distinguishing labels; the fraction of such pairs that realize a given routing matrix~$\Jb$ is precisely~$w_{\Jb}$.

We call the family of distributions $\{P_\nb(\cb)\}$, parametrized by the input composition~$\nb$ for fixed~$\Ub$ and~$m$, the \emph{quantum multinomial distribution}. It is a discrete probability distribution on compositions of~$m$, defined entirely in terms of multinomial coefficients, unitary matrix elements, and a combinatorial sum over the transportation polytope. No Hilbert space formalism is needed to state or evaluate the formula, though the derivation uses the standard Fock space description. The distribution is, strictly speaking, a classical object: a probability mass function defined by combinatorial and linear-algebraic data; the adjective ``quantum'' refers to the physical phenomenon it describes, the interference of identical bosons, rather than to the formalism in which it is expressed. The underlying algebra is an exact reorganization of the permanent of the scattering submatrix; what is new is the identification of the multivariate hypergeometric weights on the transportation polytope and the resulting characterization of quantum interference as the difference between coherent and incoherent averaging under the same combinatorial distribution.

The classical multinomial distribution is recovered when all photons enter through a single port, say $\nb = (m, 0, \ldots, 0)$: there is then exactly one routing matrix per output composition, the coherent sum has a single term, and $P(\cb \mid \nb) = \binom{m}{\cb}\prod_{j=1}^k \abs{U_{1j}}^{2c_j}$, where each photon independently selects an output port. As photons are redistributed among input ports, additional routing matrices contribute, their amplitudes interfere, and the output distribution departs from the classical multinomial. The input composition~$\nb$ thus controls the degree of non-classicality: maximal imbalance gives the classical limit, while a balanced input ($\nb = (1,1,\ldots,1)$ for $m = k$) yields the maximum number of interfering routing matrices.

The departure from classical statistics admits a sharp characterization. For \emph{distinguishable} particles, where each labeling gives rise to an independent detection event, the output probability is $P_{\mathrm{cl}}(\cb \mid \nb) = \binom{m}{\cb}\, \sum_{\Jb} w_{\Jb}\,\abs{a_{\Jb}}^2$; for \emph{identical bosons}, $P(\cb \mid \nb) = \binom{m}{\nb}\binom{m}{\cb}\, \abs{\sum_{\Jb} w_{\Jb}\, a_{\Jb}}^2$, where both sums are over routing classes with the same hypergeometric weights. The quantum-to-classical ratio is therefore
\begin{equation}\label{eq:qc-ratio}
\frac{P(\cb \mid \nb)}{P_{\mathrm{cl}}(\cb \mid \nb)} \;=\; \binom{m}{\nb} \cdot \frac{\abs{\sum_{\Jb \in \mathcal{J}} w_{\Jb}\, a_{\Jb}}^2}{\sum_{\Jb \in \mathcal{J}} w_{\Jb}\,\abs{a_{\Jb}}^2}.
\end{equation}
The prefactor $\binom{m}{\nb}$ is the bosonic enhancement from summing over input labelings; the second factor is bounded above by one via Jensen's inequality, with equality if and only if all amplitudes $a_{\Jb}$ with $w_{\Jb} > 0$ are equal.

The paper is structured as follows. In Sect.~\ref{sec:two-port}, we derive the \emph{quantum binomial family} for the two-port case (the beam splitter): the amplitude sum is reorganized into routing classes weighted by the hypergeometric distribution, and the resulting one-parameter family interpolates between the classical binomial ($n = 0$ or $n = m$) and maximally interfering statistics ($n \approx m/2$). The quantum binomial distribution is shown to be the square of an orthonormal Krawtchouk function, connecting the formulation to the theory of discrete orthogonal polynomials. The probability generating function is computed in closed form and shown not to factor, unlike the classical counterpart for distinguishable particles; the $r$-th quantum factorial moment carries a squared binomial coefficient $\binom{r}{j}^2$ in place of the classical $\binom{r}{j}$, yielding a variance that grows quadratically rather than linearly in~$m$.

In Sect.~\ref{sec:multiport}, we derive the \emph{quantum multinomial distribution} for general $k$-port interferometers: routing matrices replace routing numbers, the multivariate hypergeometric distribution replaces the univariate one, and the Chu--Vandermonde identity provides the normalization. The genuinely complex interference that arises for $k \geq 3$ from the irreducible phases of the unitary matrix is examined. In Sect.~\ref{sec:classical}, we develop the classical comparison: the Jensen characterization, the quantum-to-classical ratio, the role of the permanent, and worked examples including the Hong--Ou--Mandel effect. We conclude with a discussion of what the formulation reveals, its limitations, and open questions in Sect.~\ref{sec:discussion}.

\section{The Quantum Binomial Distribution}\label{sec:two-port}

We derive the quantum binomial family, the specialization of~\eqref{eq:quantum-multinomial} to two ports, for a lossless beam splitter. The derivation makes the combinatorial mechanism fully explicit before the notational overhead of the general $k$-port case.

\subsection{Setup}\label{sec:two-port-setup}

Consider a lossless beam splitter with transmittance~$T$ and reflectance~$R = 1 - T$, with amplitude parameters $t = \sqrt{T}$ and $r = \sqrt{R}$. The input--output relation for the creation operators is
\begin{equation}\label{eq:bs-transform}
\hat{a}^\dagger \;\to\; t\,\hat{c}^\dagger + ir\,\hat{d}^\dagger, \qquad
\hat{b}^\dagger \;\to\; ir\,\hat{c}^\dagger + t\,\hat{d}^\dagger,
\end{equation}
where $\hat{a}^\dagger$, $\hat{b}^\dagger$ are the input-port creation operators and $\hat{c}^\dagger$, $\hat{d}^\dagger$ the output-port ones. A Fock state $\ket{n, m-n}$ with $n$~photons at input port~A and $m - n$ at port~B produces the output state by expanding $(\hat{a}^\dagger)^n (\hat{b}^\dagger)^{m-n}\ket{0}$ via~\eqref{eq:bs-transform} and collecting terms with $c$~photons at output port~C and $m - c$ at port~D. We seek the probability $P(c \mid n)$ of the output configuration~$c$ given input partition~$n$. Although the derivation begins from the standard Hilbert space description, the resulting probability will depend only on combinatorial quantities ($m$, $n$, $c$, $T$, $R$) and can be stated and evaluated without reference to operators or state vectors.

\subsection{The amplitude sum and routing classes}\label{sec:routing-classes}

Each of the $n$ photons entering port~A either transmits to~C (with amplitude~$t$) or reflects to~D (with amplitude~$ir$); similarly, each of the $m - n$ photons entering port~B either reflects to~C (with amplitude~$ir$) or transmits to~D (with amplitude~$t$). The number of port-A photons that exit through~C, denoted by~$j$, determines the routing completely: $j$~photons take the path $A \to C$, $n - j$ take $A \to D$, $c - j$ take $B \to C$, and $m - n - c + j$ take $B \to D$. We call~$j$ the \emph{routing number}; it ranges from $\max(0, n + c - m)$ to $\min(n, c)$.

For a fixed routing number~$j$, the amplitude contributed by a single labeled assignment of photons to paths is
\begin{equation}\label{eq:single-amplitude}
a_j \;=\; (-1)^j\, t^{m - n - c + 2j}\, r^{n + c - 2j}.
\end{equation}
The sign $(-1)^j$ arises as follows. Each reflection contributes a factor of~$i$ to the amplitude. A routing with routing number~$j$ involves $n + c - 2j$ reflections (namely $n - j$ at port~A and $c - j$ at port~B), giving a phase factor $i^{n+c-2j} = i^{n+c}\cdot(-1)^j$. The common phase $i^{n+c}$, independent of~$j$, factors out of the sum and has unit modulus; it drops out when the squared modulus is taken. What remains is the $j$-dependent sign alternation $(-1)^j$ absorbed into~$a_j$.

The number of distinct labeled routings that realize a given routing number~$j$ is the multiplicity
\begin{equation}\label{eq:multiplicity}
\mu_j \;=\; \binom{n}{j}\binom{m-n}{c-j},
\end{equation}
counting the $\binom{n}{j}$ ways to choose which port-A photons transmit to~C and the $\binom{m-n}{c-j}$ ways to choose which port-B photons reflect to~C. The total amplitude is therefore the sum $\sum_{j=0}^{\min(n,c)} \mu_j\, a_j$, and the probability is obtained by squaring and dividing by the input normalization:
\begin{equation}\label{eq:form-I}
P(c \mid n) \;=\; \frac{c!\, (m-c)!}{n!\,(m-n)!} \left|\sum_{j=0}^{\min(n,c)} \mu_j\, a_j\right|^2.
\end{equation}
This is the direct amplitude sum, equivalent to the elementary form given by Campos, Saleh, and Teich~\cite{campos1989}.

\subsection{Hypergeometric weights and the main formula}\label{sec:two-port-formula}

The multiplicities~$\mu_j$ satisfy the Vandermonde identity
\begin{equation}\label{eq:vandermonde}
\sum_{j=0}^{\min(n,c)} \binom{n}{j}\binom{m-n}{c-j} \;=\; \binom{m}{c}.
\end{equation}
We define normalized weights
\begin{equation}\label{eq:hyp-weights}
w_j \;=\; \frac{\mu_j}{\binom{m}{c}} \;=\; \frac{\binom{n}{j}\binom{m-n}{c-j}}{\binom{m}{c}}, \qquad \sum_{j=0}^{\min(n,c)} w_j = 1.
\end{equation}
These are precisely the probability mass function of the hypergeometric distribution $\mathrm{Hyp}(n, c, m)$: the probability that, when drawing $c$~items without replacement from a population of~$m$ containing $n$~marked items, exactly~$j$ are marked. The normalization $\sum_{j=0}^{\min(n,c)} w_j = 1$ is the Vandermonde identity~\eqref{eq:vandermonde} in probabilistic form.

Substituting $\mu_j = w_j \binom{m}{c}$ into~\eqref{eq:form-I} gives a factor $\binom{m}{c}^2$ from the weights; combining with the factorial prefactor,
\begin{equation}\label{eq:prefactor-identity}
\frac{c!\,(m-c)!}{n!(m-n)!} \cdot \binom{m}{c}^2 \;=\; \frac{m!}{n!(m-n)!}\cdot\frac{m!}{c!\,(m-c)!} \;=\; \binom{m}{n}\binom{m}{c},
\end{equation}
we obtain the main formula for the two-port case:
\begin{equation}\label{eq:two-port-main}
P(c \mid n) \;=\; \binom{m}{n}\binom{m}{c} \abs*{\sum_{j=0}^{\min(n,c)} w_j\, a_j}^2.
\end{equation}
This is the specialization of~\eqref{eq:quantum-multinomial} to $k = 2$, with the routing number~$j$ replacing the routing matrix~$\Jb$, and the hypergeometric distribution replacing the multivariate hypergeometric. The prefactor $\binom{m}{n}\binom{m}{c}$ counts (input labeling, output labeling) pairs: the $\binom{m}{n}$ ways to choose which $n$ of the $m$ labels go to input port~A, times the $\binom{m}{c}$ ways to assign labels to the $m$ output photons. The fraction of such pairs that realize routing number~$j$ is precisely~$w_j$.

The inner sum $\sum_j w_j a_j$ is the expected amplitude under the hypergeometric distribution: a weighted coherent superposition of the single-path amplitudes~$a_j$, with weights determined by the combinatorial degeneracy of each routing class. The probability is then the number of labeling pairs times the squared modulus of the average amplitude per pair.

\subsection{The quantum binomial family}\label{sec:quantum-binomial}

For fixed beam splitter parameters $T$ and~$R$ and total photon number~$m$, the formula~\eqref{eq:two-port-main} defines a family of distributions $\{P_n\}_{n=0}^{m}$ on $\{0, 1, \ldots, m\}$, parametrized by the input partition number~$n$. We call this the \emph{quantum binomial} family. It is a one-parameter interpolation between two classical binomial distributions at the boundary ($n = 0$ and $n = m$), with quantum interference governing the interior.

At the boundary values $n = 0$ and $n = m$, all photons enter through a single port. The hypergeometric distribution degenerates to a point mass (only one value of~$j$ is allowed), the coherent sum reduces to a single term, and~\eqref{eq:two-port-main} yields the classical binomial distribution:
\begin{equation}\label{eq:classical-binomial}
P_0(c) \;=\; \binom{m}{c}\, T^{m-c}\, R^{c}, \qquad P_m(c) \;=\; \binom{m}{c}\, T^{c}\, R^{m-c}.
\end{equation}
No interference occurs, and each photon independently selects an output port.

For interior values $0 < n < m$, multiple routing numbers contribute to the coherent sum. The number of active terms is $\min(n, c, m-n, m-c) + 1$, maximized near $n \approx m/2$ and $c \approx m/2$. Each additional active routing number is an additional interfering pathway, and the sign alternation in the amplitudes~\eqref{eq:single-amplitude} produces oscillatory departures from the binomial envelope, including exact zeros at certain output values.

The input partition number~$n$ thus controls the degree of non-classicality: maximal imbalance ($n = 0$ or $n = m$) gives the classical binomial, while balanced input ($n \approx m/2$) gives maximal interference. Table~\ref{tab:m3} displays the complete family for $m = 3$ photons.

\begin{table}[htb]
\centering
\caption{The quantum binomial family $P_n(c)$ for $m = 3$ photons. The boundary rows ($n = 0$ and $n = 3$) are the classical binomial distributions; the interior rows show the effect of interference. Numerical values at the balanced beam splitter $T = R = \tfrac{1}{2}$ are given below.}
\label{tab:m3}
\begin{tabular}{c cccc}
\toprule
& \multicolumn{4}{c}{$P_n(c)$} \\
\cmidrule(l){2-5}
$n$ & $c = 0$ & $c = 1$ & $c = 2$ & $c = 3$ \\
\midrule
$0$ & $T^3$ & $3T^2R$ & $3TR^2$ & $R^3$ \\[3pt]
$1$ & $3T^2 R$ & $T(2R{-}T)^2$ & $R(R{-}2T)^2$ & $3TR^2$ \\[3pt]
$2$ & $3TR^2$ & $R(R{-}2T)^2$ & $T(2R{-}T)^2$ & $3T^2 R$ \\[3pt]
$3$ & $R^3$ & $3TR^2$ & $3T^2R$ & $T^3$ \\
\midrule
& \multicolumn{4}{c}{Numerical values at $T = R = \tfrac{1}{2}$} \\
\cmidrule(l){2-5}
$0$ & $\tfrac{1}{8}$ & $\tfrac{3}{8}$ & $\tfrac{3}{8}$ & $\tfrac{1}{8}$ \\[2pt]
$1$ & $\tfrac{3}{8}$ & $\tfrac{1}{8}$ & $\tfrac{1}{8}$ & $\tfrac{3}{8}$ \\[2pt]
$2$ & $\tfrac{3}{8}$ & $\tfrac{1}{8}$ & $\tfrac{1}{8}$ & $\tfrac{3}{8}$ \\[2pt]
$3$ & $\tfrac{1}{8}$ & $\tfrac{3}{8}$ & $\tfrac{3}{8}$ & $\tfrac{1}{8}$ \\
\bottomrule
\end{tabular}
\end{table}

The boundary rows are the classical binomial distributions $\mathrm{Bin}(3, T)$ and $\mathrm{Bin}(3, R)$: no interference, each photon selecting an output port independently. In the interior rows, the entries where a single routing class contributes ($c = 0$ or $c = 3$) remain simple monomials, while the entries where two routing classes interfere ($c = 1$ and $c = 2$) acquire squared-linear factors. The interference zeros are visible: $P_1(2) = R(R - 2T)^2$ vanishes at $T = 1/3$ (where $R = 2T$), and $P_1(1) = T(2R - T)^2$ vanishes at $T = 2/3$ (where $2R = T$), in each case from exact cancellation between the two routing-class amplitudes.

At the balanced beam splitter ($T = R = 1/2$), the interior distributions exhibit bunching: photons are more likely to exit through the same port ($P_1(0) = P_1(3) = 3/8$) than to split ($P_1(1) = P_1(2) = 1/8$), whereas the classical distribution assigns equal weight $3/8$ to the split outputs. This is the multi-photon manifestation of the Hong--Ou--Mandel effect~\cite{hong1987}: destructive interference between routing classes suppresses the split outputs. The table also displays the symmetry $P_n(c) = P_{m-n}(m-c)$, corresponding to the simultaneous exchange of input ports ($n \leftrightarrow m-n$) and output ports ($c \leftrightarrow m-c$); this follows from the beam splitter convention~\eqref{eq:bs-transform} having $U_{11} = U_{22} = t$, so that swapping both port labels leaves the scattering matrix invariant.

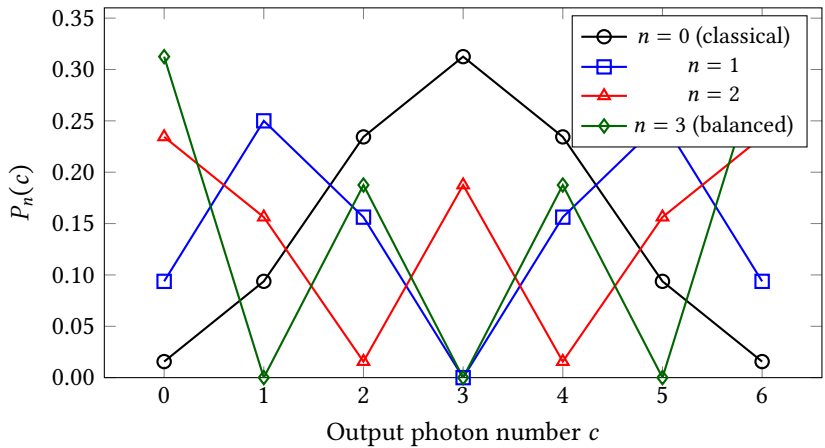
\begin{figure}[!htbp]
\centering
\begin{tikzpicture}
\begin{axis}[
  width=\figwidth,
  height=\figheight,
  xlabel={Output photon number $c$},
  ylabel={$P_n(c)$},
  xtick={0,1,2,3,4,5,6},
  ymin=0, ymax=0.36,
  ytick={0.00,0.05,0.10,0.15,0.20,0.25,0.30,0.35},
  yticklabel style={/pgf/number format/.cd, fixed, fixed zerofill, precision=2},
  legend style={at={(0.98,0.98)}, anchor=north east, font=\figlegendfont},
  every axis plot/.append style={thick, mark size=\figmarksize},
]
\addplot[black, mark=o] coordinates {
  (0, 0.015625) (1, 0.09375) (2, 0.234375) (3, 0.3125) (4, 0.234375) (5, 0.09375) (6, 0.015625)
};
\addlegendentry{$n = 0$ (classical)}
\addplot[blue, mark=square] coordinates {
  (0, 0.09375) (1, 0.25) (2, 0.15625) (3, 0) (4, 0.15625) (5, 0.25) (6, 0.09375)
};
\addlegendentry{$n = 1$}
\addplot[red, mark=triangle] coordinates {
  (0, 0.234375) (1, 0.15625) (2, 0.015625) (3, 0.1875) (4, 0.015625) (5, 0.15625) (6, 0.234375)
};
\addlegendentry{$n = 2$}
\addplot[black!60!green, mark=diamond] coordinates {
  (0, 0.3125) (1, 0) (2, 0.1875) (3, 0) (4, 0.1875) (5, 0) (6, 0.3125)
};
\addlegendentry{$n = 3$ (balanced)}
\end{axis}
\end{tikzpicture}
\caption{The quantum binomial family $P_n(c)$ for $m = 6$ photons at a balanced beam splitter ($T = R = 1/2$), for input partitions $n = 0, 1, 2, 3$. The boundary case $n = 0$ is the classical binomial $\mathrm{Bin}(6, 1/2)$. As~$n$ increases, interference produces oscillations and exact zeros; the balanced input $n = 3$ suppresses all odd output values.}
\label{fig:m6}
\end{figure}

Figure~\ref{fig:m6} shows the quantum binomial family for $m = 6$. As the input partition~$n$ moves from the boundary ($n = 0$, classical binomial) toward the center ($n = 3$, maximally balanced), the distribution develops increasingly deep oscillations: $P_1$ has one zero ($c = 3$), $P_2$ has no exact zeros but near-zeros at $c = 2$ and $c = 4$, and $P_3$ has three zeros (all odd~$c$). The $n = 3$ suppression at odd~$c$ follows from $(1 - x^2)^3 = (1-x)^3(1+x)^3$ having no odd-power terms.

\subsection{Comparison with standard formulations}\label{sec:two-port-comparison}

For reference, the same probability can be expressed via Jacobi polynomials as
\begin{equation}\label{eq:jacobi-form}
P(c \mid n) \;=\; \frac{\binom{m}{c}\,T^{m-c}\,R^{c}}{\binom{m}{n}\,(TR)^n}\,\parens{P_n^{(c-n,\,m-c-n)}(T - R)}^2,
\end{equation}
where $P_n^{(\alpha,\beta)}$ denotes the Jacobi polynomial~\cite{campos1989}, or via Wigner $d$-matrix elements~\cite{zare1988} as
\begin{equation}\label{eq:wigner-form}
P(c \mid n) \;=\; \parens{d^{\,m/2}_{n - m/2,\;c - m/2}(\theta)}^2, \qquad T = \cos^2\frac{\theta}{2},\quad R = \sin^2\frac{\theta}{2}.
\end{equation}
In the Jacobi form, the numerator $\binom{m}{c}T^{m-c} R^c$ is the classical binomial output probability while the denominator $\binom{m}{n}(TR)^n$ normalizes for the input partition; the form is compact and well suited to asymptotic analysis. The Wigner form connects to the representation theory of SU(2). However, neither formulation isolates the hypergeometric weights as a separate object, makes the number of interfering routing classes directly readable, or reveals the mechanism by which the input partition~$n$ controls the departure from classical statistics. These are the features that the reformulation~\eqref{eq:two-port-main} is designed to make transparent.

\subsection{The Krawtchouk structure}\label{sec:krawtchouk}

We now show that the quantum binomial distribution is the square of an orthonormal Krawtchouk function. Define
\begin{equation}\label{eq:gn-def}
g_n(c) \;=\; [s^n]\, (1 + Ts)^c\, (1 - Rs)^{m-c},
\end{equation}
where $[s^n]$ denotes the coefficient of~$s^n$. The generating function $(1+Ts)^c(1-Rs)^{m-c}$ is the standard one for the Krawtchouk polynomials $K_n(c; R, m)$, orthogonal with respect to the binomial distribution $\mathrm{Bin}(m, R)$:
\begin{equation}\label{eq:krawtchouk-gf}
\sum_{n=0}^{m} K_n(c)\, \frac{s^n}{n!} \;=\; (1 + Ts)^c\, (1 - Rs)^{m-c},
\end{equation}
so $g_n(c) = K_n(c)/n!$. Orthogonality follows by summing the product of two generating functions against $\mathrm{Bin}(m, R)$:
\[
\sum_{c=0}^{m} \binom{m}{c} R^c\, T^{m-c}\, (1{+}Ts)^c(1{-}Rs)^{m-c}\,(1{+}Tu)^c(1{-}Ru)^{m-c} \;=\; \bigl[R(1{+}Ts)(1{+}Tu) + T(1{-}Rs)(1{-}Ru)\bigr]^m
\]
by the binomial theorem. The bracket simplifies to $1 + TRsu$ (cross terms cancel by $T + R = 1$), so the full sum is $(1 + TRsu)^m$. Extracting the coefficient of $s^n u^\ell$ gives
\begin{equation}\label{eq:krawtchouk-orthogonality}
\sum_{c=0}^{m} \binom{m}{c}\, R^c\, T^{m-c}\, g_n(c)\, g_\ell(c) \;=\; \binom{m}{n}(TR)^n\, \delta_{n\ell}.
\end{equation}
The inner sum $\sum_j w_j a_j$ in~\eqref{eq:two-port-main} can be related to~$g_n$ via the identity $\binom{n}{j}\binom{m-n}{c-j} = \frac{\binom{m}{c}}{\binom{m}{n}}\binom{c}{j}\binom{m-c}{n-j}$, which converts the hypergeometric weights into generating-function coefficients; the result is $P_n(c) = P_0(c)\,g_n(c)^2/h_n$, where $P_0(c) = \binom{m}{c}\,T^{m-c}R^c$ is the classical binomial~\eqref{eq:classical-binomial} and $h_n = \binom{m}{n}(TR)^n$ is the squared norm in~\eqref{eq:krawtchouk-orthogonality}. Defining the orthonormal Krawtchouk function $\psi_n(c) = g_n(c)\sqrt{P_0(c)/h_n}$ gives
\begin{equation}\label{eq:squared-krawtchouk}
P_n(c) \;=\; \psi_n(c)^2.
\end{equation}
Each quantum binomial probability is the square of an orthonormal Krawtchouk function. The identity is equivalent to the known connection between the Wigner $d$-matrix and Krawtchouk polynomials~\cite{koornwinder1982}, which underlies the fractional quantum Krawtchouk transform realized experimentally at a beam splitter~\cite{stobinska2019}; the generating-function route makes the orthogonality transparent.

The identification has two further consequences. First, the normalization $\sum_{c=0}^m P_n(c) = 1$ is equivalent to the completeness (dual orthogonality) of the Krawtchouk basis, a consistency check that links unitarity of the beam splitter to a classical polynomial identity. Second, the orthogonal polynomials of the quantum binomial distribution $P_n$ for fixed~$n$ are Christoffel modifications of the Krawtchouk polynomials: $P_n$ has the density $K_n(c)^2 / (n!)^2 h_n$ with respect to $\mathrm{Bin}(m, R)$, and the recurrence coefficients of the modified polynomials can be computed from the Krawtchouk recurrence by a Darboux transformation. For $k \geq 3$, the amplitudes $a_{\Jb}$ are genuinely complex and $P_\nb(\cb) = |\sum_{\Jb} w_{\Jb} a_{\Jb}|^2$ is no longer the square of a single real function; the Krawtchouk structure does not extend directly.

For fixed~$n$ and $m \to \infty$, the Krawtchouk generating function converges to the Hermite generating function: substituting $c = mR + x\sqrt{mTR}$ and $s = \tau/\sqrt{mTR}$ into~\eqref{eq:krawtchouk-gf} and letting $m \to \infty$ gives $(1{+}Ts)^c(1{-}Rs)^{m-c} \to e^{x\tau - \tau^2/2} = \sum_{n=0}^{\infty} \mathrm{He}_n(x)\,\tau^n/n!$, where $\mathrm{He}_n$ is the probabilist's Hermite polynomial. The quantum binomial distribution, centered and scaled, therefore converges to the squared eigenfunction of the quantum harmonic oscillator:
\begin{equation}\label{eq:krawtchouk-hermite}
P_n(c)\;\longrightarrow\; \frac{\mathrm{He}_n(x)^2}{n!\,\sqrt{2\pi}}\,e^{-x^2/2}\,dx \;=\; \abs{\phi_n(x)}^2\,dx.
\end{equation}
The classical binomial ($n = 0$) gives the Gaussian $|\phi_0|^2 = e^{-x^2/2}/\sqrt{2\pi}$; each additional photon redistributed from a single input port adds one node to the limiting density, via the corresponding Hermite polynomial. The distribution is therefore asymptotically Gaussian only at the classical boundaries $n = 0$ and $n = m$; for $n \geq 1$, the $n$~zeros of the Krawtchouk polynomial persist in the limit as nodes of $\mathrm{He}_n$, and no central limit theorem applies. The Shannon entropy of $P_n$ is accordingly determined, for large~$m$, by the differential entropy of~$|\phi_n|^2$, a quantity that lacks a closed-form expression even in the continuous limit.

\subsection{Generating function and moments}\label{sec:moments}

The probability generating function of the quantum binomial distribution,
$G_n(s) = \sum_{c=0}^m P_n(c)\, s^c$,
can be computed in closed form from the Krawtchouk generating function. Since $P_n(c) = P_0(c)\,g_n(c)^2/h_n$ with $P_0(c) = \binom{m}{c}T^{m-c}R^c$ and $g_n(c) = [u^n](1+Tu)^c(1-Ru)^{m-c}$, the sum over~$c$ can be evaluated by introducing two auxiliary variables $u$ and~$v$ for the two copies of the generating function and applying the binomial theorem to collapse the sum. The result is
\begin{equation}\label{eq:pgf}
G_n(s) = \frac{1}{h_n}\,[u^n v^n]\,\bigl[T(1{-}Ru)(1{-}Rv) + sR(1{+}Tu)(1{+}Tv)\bigr]^m,
\end{equation}
where $[u^n v^n]$ denotes the coefficient of $u^n v^n$. At $s = 1$, the bracket reduces to $1 + TRuv$, and the extraction gives $[u^n v^n](1 + TRuv)^m = \binom{m}{n}(TR)^n = h_n$, recovering the normalization $G_n(1) = 1$.

The PGF~\eqref{eq:pgf} does not factor into single-variable terms, in contrast to the classical PGF $G_n^{\mathrm{cl}}(s) = (R + sT)^n\,(T + sR)^{m-n}$ for distinguishable particles, which is a product of $m$ independent Bernoulli contributions. The failure of factorization reflects the correlations introduced by bosonic symmetrization: the output photons are not independent, even conditionally on the input partition.

The bracket in~\eqref{eq:pgf}, which we denote $F(s,u,v)$, is linear in~$s$. Repeated differentiation of $F^m$ therefore takes a simple form: since $\partial F/\partial s = R(1{+}Tu)(1{+}Tv)$ is independent of~$s$,
\[
\frac{\partial^r}{\partial s^r} F^m \bigg|_{s=1} = m^{(r)}\,\bigl[R(1{+}Tu)(1{+}Tv)\bigr]^r\,(1 + TRuv)^{m-r},
\]
where $m^{(r)} = m(m{-}1)\cdots(m{-}r{+}1)$. Extracting $[u^n v^n]$ from the product $(1{+}Tu)^r(1{+}Tv)^r(1{+}TRuv)^{m-r}$ requires the exponents of~$u$ and~$v$ to match, forcing equal indices in the two Krawtchouk copies; using $m^{(r)}\binom{m-r}{n-j}/\binom{m}{n} = n^{(j)}(m{-}n)^{(r-j)}$, the factorial moments are
\begin{equation}\label{eq:factorial-moments}
\mathbb{E}[c^{(r)} \mid n] = \sum_{j=0}^{r} \binom{r}{j}^{\!2}\, T^j\, R^{r-j}\, n^{(j)}\, (m{-}n)^{(r-j)},
\end{equation}
where $c^{(r)} = c(c{-}1)\cdots(c{-}r{+}1)$ and $n^{(j)} = n(n{-}1)\cdots(n{-}j{+}1)$ are falling factorials. The classical counterpart, computed from the factored PGF, is
\begin{equation}\label{eq:factorial-moments-classical}
\mathbb{E}_{\mathrm{cl}}[c^{(r)} \mid n] = \sum_{j=0}^{r} \binom{r}{j}\, T^j\, R^{r-j}\, n^{(j)}\, (m{-}n)^{(r-j)}.
\end{equation}
The two formulas are identical in structure; the sole difference is $\binom{r}{j}^2$ in the quantum case versus $\binom{r}{j}$ in the classical. The squared coefficient is the moment-level fingerprint of the coherent-versus-incoherent distinction that runs through the paper. Computing the transition probability $P_n(c) = \abs{\text{amplitude}}^2$ involves two copies of the amplitude expansion, one from the ket and one from the bra, and the Fock-state constraint forces their multinomial indices to agree; in the PGF~\eqref{eq:pgf}, the two copies are encoded in the auxiliary variables $u$ and~$v$, and extracting $[u^n v^n]$ produces a factor $\binom{r}{j}$ from each copy, hence $\binom{r}{j}^2$. For distinguishable particles, the squared modulus is taken before the combinatorial expansion---each particle scatters independently---so only one copy of the multinomial structure survives.

For $r = 1$, both formulas give
\begin{equation}\label{eq:mean}
\mathbb{E}[c \mid n] = nT + (m{-}n)R:
\end{equation}
the quantum and classical means coincide. For $r = 2$, the quantum second factorial moment exceeds the classical one by $2TR\,n(m{-}n)$, giving
\begin{equation}\label{eq:variance}
\operatorname{Var}[c \mid n] = TR\bigl[m + 2n(m{-}n)\bigr],
\end{equation}
compared with $\operatorname{Var}_{\mathrm{cl}}[c \mid n] = mTR$; the mean and variance were first obtained by Campos, Saleh, and Teich~\cite{campos1989} via the SU(2) representation. The general factorial moment formula~\eqref{eq:factorial-moments}, and with it the $\binom{r}{j}^2$ characterization of the quantum departure, appear to be new. For Gaussian input states, photon-number moments and cumulants have recently been expressed in closed form via the loop Hafnian and a related matrix function~\cite{cardin2024}; formulas~\eqref{eq:factorial-moments}--\eqref{eq:fourth-cumulant} provide the Fock-state counterpart. The quantum excess variance $2TR\,n(m{-}n)$ vanishes at the classical boundaries $n = 0$ and $n = m$ and is maximized at $n = \lfloor m/2 \rfloor$, where the ratio $\operatorname{Var}/\operatorname{Var}_{\mathrm{cl}} = 1 + 2n(m{-}n)/m$ reaches $(m+2)/2$ for even~$m$. The quantum variance thus grows quadratically in~$m$ at balanced input, while the classical variance grows linearly.

The cumulants sharpen this comparison. Define the quantum and classical factorial-moment differences $\Delta_r = \mathbb{E}[c^{(r)} \mid n] - \mathbb{E}_{\mathrm{cl}}[c^{(r)} \mid n]$, so that $\Delta_1 = 0$ and $\Delta_2 = 2TR\,n(m{-}n)$. For the third cumulant, the conversion from factorial moments gives $\kappa_{3,Q} - \kappa_{3,\mathrm{cl}} = \Delta_3 + 3\Delta_2(1 - \mu)$ where $\mu = \mathbb{E}[c \mid n]$. A direct computation from~\eqref{eq:factorial-moments} yields $\Delta_3 = 6TR\,n(m{-}n)\bigl[T(n{-}1) + R(m{-}n{-}1)\bigr]$; since $T(n{-}1) + R(m{-}n{-}1) = \mu - 1$, this equals $3\Delta_2(\mu - 1)$ exactly. The two terms cancel, $3\Delta_2(\mu{-}1) + 3\Delta_2(1{-}\mu) = 0$, giving
\begin{equation}\label{eq:third-cumulant}
\kappa_{3,Q} = \kappa_{3,\mathrm{cl}} = TR(R - T)(2n - m),
\end{equation}
so the third cumulant (skewness) is unchanged by bosonic interference: the quantum redistribution of probability is symmetric about the mean to third order. This is consistent with the Krawtchouk-Hermite limit~\eqref{eq:krawtchouk-hermite}, since the third Hermite moment of~$|\phi_n|^2$ coincides with that of the Gaussian~$|\phi_0|^2$ for all~$n$.

The fourth cumulant, by contrast, differs. From~\eqref{eq:factorial-moments} and the standard factorial-moment-to-cumulant conversion,
\begin{equation}\label{eq:fourth-cumulant}
\kappa_{4,Q} - \kappa_{4,\mathrm{cl}} = 2TR\,n(m{-}n)\bigl[1 - 3\sigma\, TR\bigr], \qquad \sigma = n(m{-}n) + m + 3.
\end{equation}
The prefactor $2TR\,n(m{-}n)$ is positive for all non-boundary inputs, so the sign is controlled by the bracket $1 - 3\sigma\,TR$. At the balanced beam splitter ($TR = 1/4$), this is $1 - 3\sigma/4 < 0$ for all $\sigma \geq 6$, and $\sigma \geq 6$ holds for every non-boundary input (with equality at $m = 2$, $n = 1$). More generally, $\kappa_{4,Q} < \kappa_{4,\mathrm{cl}}$ whenever $T(1{-}T) > 1/(3\sigma)$, that is, whenever $T$ lies in the interval
\[
\tfrac{1}{2}\bigl(1 - \sqrt{1 - 4/(3\sigma)}\bigr) \;<\; T \;<\; \tfrac{1}{2}\bigl(1 + \sqrt{1 - 4/(3\sigma)}\bigr).
\]
For $m = 2$, $n = 1$ ($\sigma = 6$), this gives $T \in (0.059, 0.941)$; for $m = 6$, $n = 3$ ($\sigma = 18$), $T \in (0.019, 0.981)$; as $n(m{-}n) \to \infty$, the interval expands to $(0, 1)$. The fourth cumulant is therefore the lowest order at which the quantum departure is visible in the cumulant sequence: the zeros of the Krawtchouk polynomial suppress the tails and reduce the kurtosis below the classical value for all but the most asymmetric beam splitters.

\section{The Quantum Multinomial Distribution}\label{sec:multiport}

We now derive the quantum multinomial distribution for a general $k$-port lossless interferometer described by a $k \times k$ unitary matrix~$\Ub$. The algebraic steps parallel those of Sect.~\ref{sec:two-port}: group the amplitude sum by routing class, normalize the multiplicities via a combinatorial identity, and factor the result into a prefactor times a squared coherent sum. The objects that replace the routing number, the hypergeometric distribution, and the Vandermonde identity are, respectively, routing matrices, the multivariate hypergeometric distribution, and the Chu--Vandermonde identity. As for $k = 2$, the derivation begins from the Hilbert space description (expanding $\prod_{i=1}^k (\sum_{j=1}^k U_{ij}\,\hat{b}_j^\dagger)^{n_i}\ket{0}$ and collecting output terms), but the result depends only on combinatorial quantities and we state it in those terms.

\subsection{Routing matrices}\label{sec:routing-matrices}

The input Fock state is specified by a composition $\nb = (n_1, \ldots, n_k)$ of~$m$, and the output by a composition $\cb = (c_1, \ldots, c_k)$ of~$m$. A \emph{routing matrix} is a non-negative integer $k \times k$ matrix $\Jb = (J_{ij})_{i,j=1}^k$ satisfying
\begin{equation}\label{eq:routing-constraints}
\sum_{j=1}^k J_{ij} = n_i \quad \text{for all } i, \qquad \sum_{i=1}^k J_{ij} = c_j \quad \text{for all } j.
\end{equation}
The entry $J_{ij}$ counts the number of photons routed from input port~$i$ to output port~$j$. In the combinatorial literature, such matrices are the integer points of the \emph{transportation polytope} with row margins~$\nb$ and column margins~$\cb$~\cite{deloera2014}.

For the two-port case ($k = 2$), a routing matrix is determined by the single entry $J_{11} = j$ (the routing number of Sect.~\ref{sec:routing-classes}), since the row and column sum constraints fix the remaining three entries. For general~$k$, the routing matrix has $(k-1)^2$ free entries.

The amplitude contributed by a single labeled assignment of photons to paths, given routing matrix~$\Jb$, is the product of the corresponding unitary matrix elements:
\begin{equation}\label{eq:amplitude-multiport}
a_{\Jb} \;=\; \prod_{i=1}^{k}\prod_{j=1}^{k} U_{ij}^{\,J_{ij}}.
\end{equation}
The multiplicity of routing matrix~$\Jb$ is the number of ways to assign the $n_i$~photons at input port~$i$ among the $k$~output ports according to the row $(J_{i1}, \ldots, J_{ik})$:
\begin{equation}\label{eq:multiplicity-multiport}
\mu_{\Jb} \;=\; \prod_{i=1}^{k} \binom{n_i}{J_{i1}, \ldots, J_{ik}},
\end{equation}
where $\binom{n_i}{J_{i1}, \ldots, J_{ik}} = n_i! / \prod_{j=1}^k J_{ij}!$ is the multinomial coefficient. The total amplitude is then $\sum_{\Jb \in \mathcal{J}} \mu_{\Jb}\, a_{\Jb}$.

\subsection{The multivariate Chu--Vandermonde identity}\label{sec:chu-vandermonde}

The multiplicities satisfy the multivariate Chu--Vandermonde identity
\begin{equation}\label{eq:chu-vandermonde}
\sum_{\Jb \in \mathcal{J}} \mu_{\Jb} \;=\; \sum_{\Jb \in \mathcal{J}} \prod_{i=1}^{k} \binom{n_i}{J_{i1}, \ldots, J_{ik}} \;=\; \binom{m}{\cb},
\end{equation}
where the sum runs over all routing matrices with row sums~$\nb$ and column sums~$\cb$. For $k = 2$ this reduces to the Vandermonde identity~\eqref{eq:vandermonde}. The identity can be proved by a counting argument: both sides count the number of ways to assign $m$~labeled items, of which $n_i$ belong to category~$i$, into $k$~bins of sizes $c_1, \ldots, c_k$~\cite[Sect.~1.2]{stanley2012}.

We define normalized weights
\begin{equation}\label{eq:weights-multiport}
w_{\Jb} \;=\; \frac{\mu_{\Jb}}{\binom{m}{\cb}}, \qquad \sum_{\Jb \in \mathcal{J}} w_{\Jb} = 1.
\end{equation}
These are the probability mass function of the \emph{multivariate hypergeometric distribution}: the probability that $m$~labeled items belonging to $k$~categories of sizes $n_1, \ldots, n_k$, dealt uniformly at random into $k$~bins of sizes $c_1, \ldots, c_k$, produce contingency table~$\Jb$.

\subsection{The main formula}\label{sec:multiport-formula}

The transition probability is $P(\cb \mid \nb) = |\perm(\Ub_S)|^2 / (\prod_{i=1}^k n_i!\,\prod_{j=1}^k c_j!)$, where $\Ub_S$ is the $m \times m$ scattering submatrix. Grouping the $m!$ terms of the permanent by routing class gives $\perm(\Ub_S) = (\prod_{j=1}^k c_j!)\,\sum_{\Jb \in \mathcal{J}} \mu_{\Jb}\, a_{\Jb}$, and therefore
\begin{equation}\label{eq:multiport-form-I}
P(\cb \mid \nb) \;=\; \frac{\prod_{j=1}^k c_j!}{\prod_{i=1}^k n_i!}\, \abs*{\sum_{\Jb \in \mathcal{J}} \mu_{\Jb}\, a_{\Jb}}^2,
\end{equation}
the multiport analogue of~\eqref{eq:form-I}. Substituting $\mu_{\Jb} = w_{\Jb}\,\binom{m}{\cb}$ and using the prefactor identity
\begin{equation}\label{eq:prefactor-identity-multiport}
\frac{\prod_{j=1}^k c_j!}{\prod_{i=1}^k n_i!} \cdot \binom{m}{\cb}^2 \;=\; \frac{m!}{\prod_{i=1}^k n_i!}\cdot\frac{m!}{\prod_{j=1}^k c_j!} \;=\; \binom{m}{\nb}\binom{m}{\cb},
\end{equation}
we obtain
\begin{equation}\label{eq:multiport-main}
P(\cb \mid \nb) \;=\; \binom{m}{\nb}\binom{m}{\cb} \abs*{\sum_{\Jb \in \mathcal{J}} w_{\Jb}\, a_{\Jb}}^2,
\end{equation}
which is~\eqref{eq:quantum-multinomial}. The interpretation carries over from the two-port case: $\binom{m}{\nb}\binom{m}{\cb}$ counts the number of (input labeling, output labeling) pairs, $w_{\Jb}$ is the fraction of such pairs that realize routing matrix~$\Jb$, and the inner sum $\sum_{\Jb} w_{\Jb}\, a_{\Jb}$ is the expected amplitude per pair under the multivariate hypergeometric distribution.

\subsection{The quantum multinomial family and the classical limit}\label{sec:qm-family}

For fixed unitary~$\Ub$ and total photon number~$m$, the formula~\eqref{eq:multiport-main} defines a family of distributions $\{P_\nb\}$ on compositions of~$m$ into~$k$ parts, parametrized by the input composition~$\nb$. We call this the \emph{quantum multinomial} family: it generalizes the classical multinomial distribution in the same way that the quantum binomial family of Sect.~\ref{sec:quantum-binomial} generalizes the classical binomial, with the input composition~$\nb$ controlling the departure from classical statistics through the number and phases of the interfering routing classes.

The classical multinomial is recovered when all photons enter through a single port, say $\nb = m\,\eb_i$ (the $i$-th standard basis vector scaled by~$m$). There is then exactly one routing matrix for each output composition~$\cb$, namely the matrix with row~$i$ equal to~$\cb$ and all other rows zero. The coherent sum has a single term, $w_{\Jb} = 1$, and
\begin{equation}\label{eq:classical-multinomial}
P_{m\,\eb_i}(\cb) \;=\; \binom{m}{\cb} \prod_{j=1}^{k} \abs{U_{ij}}^{2c_j},
\end{equation}
which is the multinomial distribution with probabilities $p_j = |U_{ij}|^2$: each photon independently selects an output port. As photons are redistributed among input ports, additional routing matrices contribute and their amplitudes interfere, producing departures from multinomial statistics that range from mild redistribution of probability to complete suppression of certain outputs.

The number of routing matrices for a given pair $(\nb, \cb)$, that is, the number of integer points in the transportation polytope with margins~$\nb$ and~$\cb$, determines how many terms participate in the coherent sum:
\begin{center}
\begin{tabular}{lcc}
\toprule
Input configuration & Routing matrices & Interference \\
\midrule
Single port: $\nb = m\,\eb_i$ & $1$ for all $\cb$ & None (classical) \\
All singly occupied: $\nb = (1,\ldots,1)$, $m = k$ & Up to $k!$ & Maximal \\
\bottomrule
\end{tabular}
\end{center}
In the last case, the routing matrices with $\cb = (1, \ldots, 1)$ are precisely the $k!$ permutation matrices, and the coherent sum is a weighted sum over all permutations of the unitary matrix elements.

\paragraph{Symmetry.} The formula~\eqref{eq:multiport-main} satisfies $P_{\Ub}(\cb \mid \nb) = P_{\Ub^T}(\nb \mid \cb)$: transposing each routing matrix~$\Jb$ exchanges row and column sums ($\nb \leftrightarrow \cb$) and replaces $U_{ij}^{\,J_{ij}}$ by $(\Ub^T)_{ji}^{\,J_{ij}}$; the identity then follows from $\perm(M) = \perm(M^T)$ applied to the scattering submatrix. For the two-port case with the symmetric beam splitter convention~\eqref{eq:bs-transform}, where $\Ub = \Ub^T$, this gives $P_n(c) = P_c(n)$: the time-reversal symmetry, swapping input and output. A separate symmetry arises from port exchange: for any permutation~$\Pi$ of the $k$~ports, $P_{\Ub}(\cb \mid \nb) = P_{\Pi \Ub \Pi^T}(\Pi\cb \mid \Pi\nb)$, with equality when $\Pi \Ub \Pi^T = \Ub$. For the beam splitter~\eqref{eq:bs-transform}, the swap $\Pi = \bigl(\begin{smallmatrix} 0 & 1 \\ 1 & 0\end{smallmatrix}\bigr)$ leaves~$\Ub$ invariant since $U_{11} = U_{22}$, yielding $P_n(c) = P_{m-n}(m-c)$ as visible in Table~\ref{tab:m3}.

\subsection{Phase structure for $k \geq 3$}\label{sec:phase-structure}

The multiport case is genuinely richer than the beam splitter, and the quantum multinomial formula makes the reason visible: the amplitudes~$a_{\Jb}$ become complex in a way that has no two-port analogue.

A $k \times k$ unitary matrix has $(k-1)^2$ physically observable parameters (after removing $2k - 1$ unobservable input and output port phases). In the Reck decomposition~\cite{reck1994}, these correspond to $k(k-1)/2$ beam splitter transmittances and $(k-1)(k-2)/2$ internal phase shifts:
\begin{equation}\label{eq:parameter-decomposition}
(k-1)^2 \;=\; \underbrace{\tfrac{k(k-1)}{2}}_{\text{mixing angles}} \;+\; \underbrace{\tfrac{(k-1)(k-2)}{2}}_{\text{irreducible phases}}.
\end{equation}

The moduli $|U_{ij}|^2$ form a doubly stochastic matrix ($k$ row sums and $k$ column sums equal to one, one constraint redundant), so the number of free moduli is also $(k-1)^2$. For $k = 2$, a single parameter~$T$ determines all moduli. For $k \geq 3$, the $k(k-1)/2$ mixing angles do not suffice: the irreducible phases also affect the moduli $|U_{ij}|^2$, and hence the classical output probabilities.

For $k = 2$, there are no irreducible phases: all amplitudes~$a_j$ are real up to a common phase factor (the factor $i^{n+c}$ of Sect.~\ref{sec:routing-classes}), and interference reduces to sign alternation. For $k \geq 3$, the irreducible phases make the amplitudes~$a_{\Jb}$ genuinely complex: two routing classes whose amplitudes merely differ in sign for a real unitary may have an arbitrary relative phase for a complex one, and the interference pattern acquires the full angular structure of the complex plane.

For $k = 3$, there is one irreducible phase, the analogue of the CP-violating Dirac phase in the CKM and PMNS mixing matrices of particle physics. When there is a single routing matrix, as in the classical limit $\nb = m\,\eb_i$, the probability $P = \binom{m}{\nb}\binom{m}{\cb}\,|a_{\Jb}|^2$ involves no interference. When several routing classes contribute, their coherent combination is highly sensitive to this phase: for a representative tritter (three mixing angles fixed, single irreducible phase varied), $P\bigl((1,1,1) \mid (1,1,1)\bigr)$, which involves six routing matrices, varies by several orders of magnitude as the phase ranges over~$[0, \pi]$; the classical probability for the same transition changes by less than~$10\%$.

\paragraph{Example: the Fourier interferometer.} The $k$-port Fourier (DFT) interferometer, $U_{ij} = \omega^{ij}/\sqrt{k}$ with $\omega = e^{2\pi i/k}$, is the multiport generalization of the balanced beam splitter. For this matrix, $|U_{ij}|^2 = 1/k$ for all $i, j$, so the classical output distribution is the symmetric multinomial $\binom{m}{\cb}\,k^{-m}$, each photon selecting an output port uniformly at random. The amplitude takes the form
\begin{equation}\label{eq:fourier-amplitude}
a_{\Jb} \;=\; k^{-m/2}\, \omega^{\sum_{i=1}^k \sum_{j=1}^k ij\, J_{ij}},
\end{equation}
where all amplitudes have the same modulus $k^{-m/2}$ and differ only in phase. The interference is purely a phase effect: the weighted sum
\begin{equation}\label{eq:fourier-sum}
\sum_{\Jb \in \mathcal{J}} w_{\Jb}\, a_{\Jb} \;=\; k^{-m/2}\sum_{\Jb \in \mathcal{J}} w_{\Jb}\, \omega^{f(\Jb)}, \qquad f(\Jb) = \textstyle\sum_{i=1}^k \sum_{j=1}^k ij\, J_{ij},
\end{equation}
is a weighted character sum over the transportation polytope. Whether this sum vanishes, and with it the transition probability, depends on the symmetry of the generating function. For the Fourier tritter ($k = 3$, $\omega = e^{2\pi i/3}$), defining $Z_i = z_1 + \omega^i\, z_2 + \omega^{2i}\, z_3$, the generating function
\begin{equation}\label{eq:tritter-gf}
G_\nb(\mathbf{z}) \;=\; \prod_{i=1}^{3} \Bigl(\frac{1}{\sqrt{3}}\sum_{j=1}^{3} \omega^{ij}\, z_j\Bigr)^{\!n_i} \;=\; 3^{-m/2}\, Z_1^{n_1}\, Z_2^{n_2}\, Z_3^{n_3}
\end{equation}
factors into DFT modes, echoing the factored beam-splitter generating function~\eqref{eq:krawtchouk-gf}. The trivial mode $Z_3 = z_1 + z_2 + z_3$ contributes only to the total photon count; the interference is carried by $Z_1$ and $Z_2$, whose product $Z_1 Z_2 = e_1^2 - 3e_2$ (with $e_r$ the elementary symmetric polynomials in $z_1, z_2, z_3$) is invariant under all permutations of the output ports. Three examples illustrate how the symmetry of~\eqref{eq:tritter-gf} controls suppression.

\paragraph{Balanced-output suppression.} For input $\nb = (2,1,0)$ and output $\cb = (1,1,1)$, there are three routing matrices with equal weights $w_{\Jb} = 1/3$ and phases $1, \omega, \omega^2$. The weighted sum $\sum_{\Jb} w_{\Jb}\, a_{\Jb} \propto 1 + \omega + \omega^2 = 0$ vanishes by the roots-of-unity identity, giving $P = 0$: a \emph{suppression law}. The cancellation is an instance of a $\mathbb{Z}_3$ selection rule: the cyclic substitution $z_j \to z_{j+1}$ sends $Z_i \to \omega^{2i}\, Z_i$ and hence $G_\nb \to \omega^{2n_1 + n_2}\, G_\nb$; at a balanced output $\cb = (d,d,d)$, the monomial $(z_1 z_2 z_3)^d$ is cyclically invariant, so $P(\cb \mid \nb) = 0$ unless $2n_1 + n_2 \equiv 0\pmod{3}$.

\paragraph{Partial cancellation.} For input $\nb = \cb = (1,1,1)$, the routing matrices are the $3! = 6$ permutation matrices, each with weight $w_{\Jb} = 1/6$ and amplitude $a_\sigma = 3^{-3/2}\, \omega^{\sum_{i=1}^3 i\cdot\sigma(i)}$. The even permutations contribute phase~$\omega^2$ and the odd permutations contribute phase~$\omega$, so
\[
\textstyle\sum_{\Jb \in \mathcal{J}} w_{\Jb}\, a_{\Jb} \;=\; \tfrac{1}{6}\cdot 3^{-3/2}\bigl(3\omega^2 + 3\omega\bigr) \;=\; \tfrac{-1}{6\sqrt{3}},
\]
using $\omega + \omega^2 = -1$. The partial cancellation gives $P(\cb \mid \nb) = 1/3$, compared with $P_{\mathrm{cl}} = 2/9$; the quantum-to-classical ratio is $3/2$, well below the Jensen bound $\binom{3}{1,1,1} = 6$. The input $\nb = (1,1,1)$ satisfies $2n_1 + n_2 = 3 \equiv 0$, consistent with the $\mathbb{Z}_3$ rule: the transition is allowed but partially suppressed.

\paragraph{Equal-pair suppression.} For $m = 4$ photons with input $\nb = (2,2,0)$, the generating function $G = 3^{-2}(Z_1 Z_2)^2 = 9^{-1}(e_1^2 - 3e_2)^2$ is invariant under all permutations of $(z_1, z_2, z_3)$, so $P(\cb \mid \nb)$ depends only on the partition type of~$\cb$. Expanding $(e_1^2 - 3e_2)^2 = e_1^4 - 6\,e_1^2 e_2 + 9\,e_2^2$ in the monomial symmetric basis, the coefficient of the partition type~$(2,1,1)$ vanishes: $12 - 30 + 18 = 0$. Therefore $P(\cb \mid \nb) = 0$ for all $\cb \in \{(2,1,1),\, (1,2,1),\, (1,1,2)\}$. The mechanism is distinct from the $\mathbb{Z}_3$ rule: it arises from the full permutation symmetry of $Z_1 Z_2$, not from its cyclic part.

The classification of all suppressed transitions for general~$k$ is an open combinatorial question; for partial results, see~\cite{tichy2010}.

\subsection{Single-mode and cross-mode moments}\label{sec:multiport-moments}

The single-mode factorial moments of the quantum multinomial extend the beam splitter formula~\eqref{eq:factorial-moments} to arbitrary~$k$. In the Heisenberg picture, $\hat{b}_j = \sum_{i=1}^k U_{ij}^* \hat{a}_i$, and the $r$-th factorial moment is $\mathbb{E}[c_j^{(r)} \mid \nb] = \langle (\hat{b}_j^\dagger)^r\,(\hat{b}_j)^r \rangle_\nb$. Expanding both powers by the multinomial theorem produces a sum over compositions~$\mathbf{q}$ of~$r$ into $k$~parts; the Fock state expectation forces the compositions from the creation and annihilation sides to match, giving
\begin{equation}\label{eq:multiport-factorial-moments}
\mathbb{E}[c_j^{(r)} \mid \nb] = \sum_{\abs{\mathbf{q}}=r} \binom{r}{\mathbf{q}}^{\!2}\, \prod_{i=1}^k p_{ij}^{q_i}\, n_i^{(q_i)},
\end{equation}
where $p_{ij} = \abs{U_{ij}}^2$ and $\binom{r}{\mathbf{q}} = r!/\prod_{i=1}^k q_i!$ is the multinomial coefficient. For $k = 2$, this reduces to~\eqref{eq:factorial-moments}. The classical counterpart for distinguishable particles has $\binom{r}{\mathbf{q}}$ in place of $\binom{r}{\mathbf{q}}^2$: in the quantum case, the coherent squaring $\abs{\text{amplitude}}^2$ generates two copies of the multinomial expansion whose compositions the Fock state forces to agree, producing the extra factor.

The formula depends only on $p_{ij} = \abs{U_{ij}}^2$, not on the phases of~$\Ub$. In particular, the mean $\mathbb{E}[c_j \mid \nb] = \sum_{i=1}^k p_{ij}\, n_i$ coincides with the classical value. For $r = 2$, the quantum excess over the classical variance $\operatorname{Var}_{\mathrm{cl}}[c_j \mid \nb] = \sum_{i=1}^k p_{ij}(1 - p_{ij})\,n_i$ is
\begin{equation}\label{eq:multiport-variance}
\operatorname{Var}[c_j \mid \nb] - \operatorname{Var}_{\mathrm{cl}}[c_j \mid \nb] = 2\!\sum_{1 \leq i < i' \leq k} p_{ij}\,p_{i'j}\,n_i\,n_{i'},
\end{equation}
which vanishes when at most one input port is occupied (the classical limit) and grows with the number of occupied input pairs. For the Fourier interferometer ($p_{ij} = 1/k$), the excess is $(m^2 - \norm{\nb}^2)/k^2$, maximized at the balanced input $\nb = (1, \ldots, 1)$. The quantum-to-classical variance ratio is
\[
\frac{\operatorname{Var}[c_j]}{\operatorname{Var}_{\mathrm{cl}}[c_j]} = 1 + \frac{m^2 - \norm{\nb}^2}{m(k-1)},
\]
which for single-photon inputs $\nb = (1, \ldots, 1)$ with $m = k$ equals~$2$, independently of~$k$. By contrast, the beam splitter with balanced input gives $(m+2)/2$, which grows without bound. Distributing photons across more ports thus moderates the variance enhancement: the multiport quantum excess remains bounded while the two-port excess scales linearly with photon number.

Cross-mode factorial moments, by contrast, involve the phases of~$\Ub$. The second factorial cross-moment of two distinct outputs $j \neq l$ is
\begin{equation}\label{eq:cross-moment}
\mathbb{E}[c_j\, c_l \mid \nb] = \sum_{i=1}^{k} p_{ij}\,p_{il}\,n_i^{(2)} + \sum_{\substack{i,i'=1 \\ i \neq i'}}^{k}\Bigl(p_{ij}\,p_{i'l} + U_{ij}\overline{U_{i'j}}\,U_{i'l}\overline{U_{il}}\Bigr) n_i\, n_{i'},
\end{equation}
where the coherence terms $U_{ij}\overline{U_{i'j}}\,U_{i'l}\overline{U_{il}}$ vanish in the classical (distinguishable-particle) case. The quantum excess covariance is therefore $\sum_{\substack{i,i'=1 \\ i \neq i'}}^{k} U_{ij}\overline{U_{i'j}}\,U_{i'l}\overline{U_{il}}\,n_i\, n_{i'}$, which depends on the phases of~$\Ub$ and can be negative: bosonic interference can strengthen the output anti-correlations beyond the classical value. For the balanced beam splitter ($k = 2$, $T = R = 1/2$) with input $\nb = (1,1)$, each coherence term evaluates to $U_{i1}\overline{U_{i'1}}\,U_{i'2}\overline{U_{i2}} = -1/4$, giving $\operatorname{Cov}_Q(c_1, c_2) = -1$ versus $\operatorname{Cov}_{\mathrm{cl}}(c_1, c_2) = -1/2$: the quantum anti-correlation is twice as strong, the covariance-level expression of Hong--Ou--Mandel bunching. Two-point output correlations of this type have been proposed as efficient statistical benchmarks for boson sampling~\cite{walschaers2016}; the formula~\eqref{eq:cross-moment} gives the closed-form expression from which such benchmarks can be evaluated for any interferometer and input configuration. For the Fourier interferometer ($U_{ij} = \omega^{ij}/\sqrt{k}$, $\omega = e^{2\pi i/k}$), the coherence terms take the form $U_{ij}\overline{U_{i'j}}\,U_{i'l}\overline{U_{il}} = k^{-2}\,\omega^{(i-i')(j-l)}$. For $j \neq l$ and $\nb = (1, \ldots, 1)$, summing over $i \neq i'$ gives $k^{-2}\bigl(\abs{\sum_{i=1}^k \omega^{i(l-j)}}^2 - k\bigr) = -1/k$, since the inner sum vanishes by the roots-of-unity identity. The quantum excess covariance is therefore $-1/k$, giving $\operatorname{Cov}_Q(c_j, c_l) = -2/k$ versus $\operatorname{Cov}_{\mathrm{cl}}(c_j, c_l) = -1/k$: the ratio is exactly~$2$, independent of~$k$. The factor-of-two enhancement of the Hong--Ou--Mandel anti-correlation thus persists for the entire Fourier family.

\subsection{Cumulants}\label{sec:multiport-cumulants}

A closed-form probability generating function analogous to~\eqref{eq:pgf} is not available for $k \geq 3$: the beam splitter PGF relies on the Krawtchouk factorization $P_n(c) = \psi_n(c)^2$, which does not extend to the multiport case where the amplitudes are genuinely complex and the transition probability is an irreducible squared modulus. Nevertheless, the cumulants can be extracted directly from~\eqref{eq:multiport-factorial-moments}. For $k = 2$, the third cumulant is invariant under bosonic interference (Sect.~\ref{sec:moments}); remarkably, this invariance breaks for $k \geq 3$. The third-cumulant difference is
\[
\kappa_{3,Q} - \kappa_{3,\mathrm{cl}} = \Delta_3 + 3\Delta_2(1 - \mu_j),
\]
where $\Delta_r = \mathbb{E}_Q[c_j^{(r)}] - \mathbb{E}_{\mathrm{cl}}[c_j^{(r)}]$ and $\mu_j = \mathbb{E}[c_j]$. For $k = 2$, the pair-sum identity $\Delta_3 = 3\Delta_2(\mu_j - 1)$ makes this vanish; for $k \geq 3$, the composition $\mathbf{q} = (1, 1, 1, 0, \ldots)$ contributes a three-body term with coefficient $\binom{3}{1,1,1}^2 - \binom{3}{1,1,1} = 30$ that has no two-port analogue and breaks the cancellation. For the Fourier interferometer with $\nb = (1, \ldots, 1)$, only $(1,1,1)$-type compositions survive (since $n_i^{(2)} = 0$), and the difference simplifies to
\[
\kappa_{3,Q} - \kappa_{3,\mathrm{cl}} = \frac{5(k-1)(k-2)}{k^2};
\]
the factor $(k - 2)$ vanishes at $k = 2$ and is positive for all $k \geq 3$. The quantum departure thus enters one cumulant order earlier for multiport interferometers than for the beam splitter: the third cumulant is the first to feel the interference when three or more ports are available.

\section{Classical Comparison and the Permanent}\label{sec:classical}

The quantum multinomial formula~\eqref{eq:multiport-main} was derived from the Hilbert space description of identical bosons. We now derive the output distribution for \emph{distinguishable} particles in the same interferometer. The two formulas share the same combinatorial ingredients (routing matrices, hypergeometric weights, amplitudes) and differ in a single respect: whether the squared modulus is taken before or after averaging over routing classes.

\subsection{Distinguishable particles}\label{sec:distinguishable}

Consider $m$ distinguishable particles, $n_i$ of which enter input port~$i$, each independently scattered by the interferometer: a particle at input~$i$ exits at output~$j$ with probability $p_{ij} = |U_{ij}|^2$. The probability that the routing matrix is~$\Jb$ is a product of independent multinomials over input ports:
\begin{equation}\label{eq:classical-routing}
\Pr(\Jb) \;=\; \prod_{i=1}^{k} \binom{n_i}{J_{i1}, \ldots, J_{ik}} \prod_{i=1}^{k}\prod_{j=1}^{k} |U_{ij}|^{2J_{ij}} \;=\; \mu_{\Jb}\, |a_{\Jb}|^2.
\end{equation}
The same routing matrices, the same multiplicities~$\mu_{\Jb}$, and the same amplitudes~$a_{\Jb}$ appear as in the quantum case; the difference is that each routing class contributes its squared amplitude~$|a_{\Jb}|^2$ independently, rather than its amplitude~$a_{\Jb}$ coherently. Summing over all routing matrices with column sums~$\cb$,
\begin{equation}\label{eq:classical-output}
P_{\mathrm{cl}}(\cb \mid \nb) \;=\; \sum_{\Jb \in \mathcal{J}} \mu_{\Jb}\, |a_{\Jb}|^2 \;=\; \binom{m}{\cb}\, \sum_{\Jb \in \mathcal{J}} w_{\Jb}\, |a_{\Jb}|^2,
\end{equation}
where the sum on the right is over routing classes with the hypergeometric weights~\eqref{eq:weights-multiport}. The classical output probability is the multinomial coefficient times the incoherent average of the squared amplitudes.

Comparing with the quantum formula $P(\cb \mid \nb) = \binom{m}{\nb}\binom{m}{\cb}\,|\sum_{\Jb} w_{\Jb}\, a_{\Jb}|^2$, we see that both involve the same weighted sum over routing classes: the quantum formula sums the amplitudes first and squares ($|\sum_{\Jb} w_{\Jb}\, a_{\Jb}|^2$, coherent summation), while the classical formula squares first and sums ($\sum_{\Jb} w_{\Jb}\,|a_{\Jb}|^2$, incoherent summation). The reason the same weights appear in both cases is that both calculations originate from a uniform sum over labeled assignments: if $\Omega$ denotes the set of all labeled routings consistent with~$(\nb, \cb)$ and each $\omega \in \Omega$ contributes amplitude $a(\omega) = a_{\Jb(\omega)}$, then $w_{\Jb}$ is the pushforward of the uniform measure on~$\Omega$ onto routing classes. The hypergeometric weights are not an additional ingredient; they are the counting measure on labeled assignments, projected onto equivalence classes.

For identical \emph{fermions}, the Pauli exclusion principle restricts inputs and outputs to collision-free configurations ($n_i, c_j \in \{0, 1\}$, hence $m \leq k$). The routing matrices are then permutation matrices with uniform weights $w_{\Jb} = 1/m!$, and the permanent is replaced by the determinant: $P_{\mathrm{ferm}}(\cb \mid \nb) = \binom{m}{\nb}\binom{m}{\cb}\,\abs{\sum_{\Jb \in \mathcal{J}} w_{\Jb}\, \mathrm{sgn}(\Jb)\, a_{\Jb}}^2$, where $\mathrm{sgn}(\Jb)$ is the signature of the permutation. The three particle statistics thus differ only in how the amplitudes are combined: coherently ($\sum w_{\Jb}\, a_{\Jb}$, bosons), coherently with signs ($\sum w_{\Jb}\, \mathrm{sgn}(\Jb)\, a_{\Jb}$, fermions), or incoherently ($\sum w_{\Jb}\, |a_{\Jb}|^2$, distinguishable). The determinant, unlike the permanent, is efficiently computable, so the fermionic case carries none of the computational complexity of the bosonic one.

\subsection{The quantum-to-classical ratio}\label{sec:qc-ratio-section}

From the formulas of the preceding subsection, the ratio stated in~\eqref{eq:qc-ratio},
\[
\frac{P(\cb \mid \nb)}{P_{\mathrm{cl}}(\cb \mid \nb)} \;=\; \binom{m}{\nb} \cdot \frac{|\sum_{\Jb \in \mathcal{J}} w_{\Jb}\, a_{\Jb}|^2}{\sum_{\Jb \in \mathcal{J}} w_{\Jb}\,|a_{\Jb}|^2},
\]
decomposes into two factors with distinct origins.

The \emph{symmetrization prefactor}~$\binom{m}{\nb}$ reflects the difference in how the two formulas treat the input configuration. For distinguishable particles, the input labeling is fixed and only the $\binom{m}{\cb}$ output arrangements contribute; for identical bosons, the $\binom{m}{\nb}$ input labelings also contribute coherently, giving a total of $\binom{m}{\nb}\binom{m}{\cb}$ labeling pairs. The ratio of labeling counts is~$\binom{m}{\nb}$. This factor depends only on the input partition, not on the unitary matrix or the output composition.

The \emph{interference factor} $|\sum_{\Jb} w_{\Jb}\, a_{\Jb}|^2 / \sum_{\Jb} w_{\Jb}\,|a_{\Jb}|^2$ is bounded between zero and one by Jensen's inequality (since $z \mapsto |z|^2$ is convex). It equals one when all amplitudes~$a_{\Jb}$ with $w_{\Jb} > 0$ are equal (no destructive interference), and vanishes when $\sum_{\Jb} w_{\Jb}\, a_{\Jb} = 0$ (complete destructive interference). The overall ratio thus satisfies $0 \leq P/P_{\mathrm{cl}} \leq \binom{m}{\nb}$.

At the classical limit $\nb = m\,\eb_i$, there is a single routing matrix for each output composition, the hypergeometric distribution is a point mass, and both factors are trivially one: $P = P_{\mathrm{cl}}$. As photons are redistributed among input ports, the symmetrization prefactor grows (more input labelings) while the interference factor can decrease (more routing classes with potentially misaligned phases). The balance between the two determines whether a given output is enhanced or suppressed relative to the classical prediction.

\paragraph{Example: the Hong--Ou--Mandel dip.} For two photons at a balanced beam splitter ($m = 2$, $\nb = (1,1)$, $T = R = 1/2$), the output $\cb = (1,1)$ has two routing classes with equal weights $w_0 = w_1 = 1/2$ and amplitudes $a_0 = 1/2$, $a_1 = -1/2$. The coherent sum vanishes by sign cancellation ($\sum_j w_j\, a_j = 0$), while the incoherent sum is $\sum_j w_j\, |a_j|^2 = 1/4$. The quantum probability is zero; the classical probability is $\binom{2}{1} \cdot 1/4 = 1/2$. The Hong--Ou--Mandel dip~\cite{hong1987} is the simplest suppression law: complete destructive interference between two equally weighted routing classes. The bunched output $\cb = (2, 0)$, by contrast, has a single routing class, giving $P/P_{\mathrm{cl}} = \binom{2}{1} = 2$: the full symmetrization bonus with no destructive interference. For a multiport suppression example involving complex phases, see the Fourier tritter case at the end of Sect.~\ref{sec:phase-structure}.

\paragraph{Example: partial interference.} For $m = 4$ photons at a balanced beam splitter with input $n = c = 2$, three routing classes contribute, with hypergeometric weights $w_0 = w_2 = 1/6$, $w_1 = 2/3$ and amplitudes $a_0 = a_2 = 1/4$, $a_1 = -1/4$. The majority class ($j = 1$, weight~$2/3$, negative amplitude) partially cancels the two minority classes ($j = 0, 2$, combined weight~$1/3$, positive amplitudes), giving $\sum_j w_j\, a_j = -1/12$ and $P_2(2) = 1/4$. The classical probability is $P_{\mathrm{cl}} = 3/8$ and the ratio $P/P_{\mathrm{cl}} = 2/3$, well below the Jensen bound~$\binom{4}{2} = 6$: a representative intermediate case between complete cancellation ($P/P_{\mathrm{cl}} = 0$, the dip) and no interference ($P/P_{\mathrm{cl}} = 1$, the classical limit).

\subsection{The permanent}\label{sec:permanent}

The standard expression for the transition probability involves the permanent of an $m \times m$ scattering submatrix~$\Ub_S$, formed by repeating row~$i$ of~$\Ub$ a total of $n_i$~times and column~$j$ a total of $c_j$~times~\cite{scheel2008}:
\begin{equation}\label{eq:permanent-formula}
P(\cb \mid \nb) \;=\; \frac{|\perm(\Ub_S)|^2}{\prod_{i=1}^k n_i!\,\prod_{j=1}^k c_j!}.
\end{equation}
The permanent sums over all $m!$ permutations of photon-label assignments. As shown in Sect.~\ref{sec:multiport-formula}, grouping by routing class gives $\perm(\Ub_S) = (\prod_{j=1}^k c_j!)\,\sum_{\Jb \in \mathcal{J}} \mu_{\Jb}\, a_{\Jb}$, so the quantum multinomial formula is an exact reorganization of the permanent, not an approximation.

The reorganization reduces the number of terms from $m!$ (one per labeled permutation) to the number of integer points in the transportation polytope with margins~$(\nb, \cb)$ (one per routing class). For $m = 10$ photons at a balanced beam splitter ($\nb = (5,5)$, $\cb = (5,5)$), this gives $6$ routing classes versus $10! = 3{,}628{,}800$ permanent terms. The reduction is possible because identical photons within the same routing class contribute the same amplitude, and their contributions collect into the multiplicity~$\mu_{\Jb}$.

The permanent of a general complex matrix is \#P-hard to compute~\cite{valiant1979}, and this intractability underlies the boson sampling proposal~\cite{aaronson2013}: the output distribution of a linear optical network with single-photon inputs cannot be efficiently sampled by a classical computer. In the quantum multinomial formulation, the source of this hardness is visible. When $m = k$ and $\nb = \cb = (1, \ldots, 1)$, every routing matrix is a permutation matrix, all multiplicities are $\mu_{\Jb} = 1$, and the coherent sum becomes
\begin{equation}\label{eq:perm-identity}
\sum_{\Jb \in \mathcal{J}} w_{\Jb}\, a_{\Jb} \;=\; \frac{1}{k!}\sum_{\sigma \in S_k} \prod_{i=1}^k U_{i,\sigma(i)} \;=\; \frac{\perm(\Ub)}{k!}.
\end{equation}
The hardness arises from the exponential proliferation of routing classes ($k!$ permutation matrices) combined with generically incommensurate complex phases, precisely the regime where Jensen's inequality is far from tight and no efficient shortcut for evaluating the coherent sum is known.

\section{Discussion and Conclusion}\label{sec:discussion}

The quantum multinomial distribution expresses multiphoton transition probabilities as a coherent average over routing matrices, weighted by the multivariate hypergeometric distribution. The same routing matrices, the same weights, and the same amplitudes appear in the classical calculation for distinguishable particles; the only difference is whether the squared modulus is taken before or after the average. That a classical combinatorial object, the hypergeometric distribution, organizes a quintessentially quantum phenomenon is not an assumption of the formulation but a consequence: regrouping the permanent's $m!$ terms by routing class produces degeneracy factors that are precisely the hypergeometric probabilities. The input composition~$\nb$ controls both the size of the coherent sum (how many routing classes contribute) and the symmetrization prefactor (how many input labelings are coherently superposed), while Jensen's inequality bounds the interference factor between complete cancellation and full constructive interference. The classical multinomial sits at the boundary of the family, where the sum has a single term and no interference is possible. Despite its name, the quantum multinomial distribution is itself a classical object: a probability mass function built from multinomial coefficients, hypergeometric weights, and products of unitary matrix elements. The adjective ``quantum'' refers to the physical phenomenon it describes, not to the formalism; that a quintessentially quantum effect can be captured by classical combinatorics is, in a sense, the point of the reformulation.

The most concrete open direction is the classification of suppression laws. The permanent formulation identifies suppression with the vanishing of a sum over $m!$ complex terms, a condition that is opaque. The quantum multinomial replaces this with $\sum_{\Jb} w_{\Jb}\, a_{\Jb} = 0$, where the weights are known and the amplitudes are products of unitary matrix elements; for the Fourier interferometer, the condition reduces to the vanishing of a character sum over the transportation polytope. Which pairs~$(\nb, \cb)$ are suppressed, and how the answer depends on the arithmetic of the unitary matrix, is a well-posed combinatorial question that the formulation is designed to make tractable. A second direction concerns the orthogonal polynomials of the quantum multinomial family. At the classical boundary ($\nb = m\,\eb_i$), the distribution is the multinomial, whose orthogonal polynomials are the Krawtchouk family. As the input composition moves into the interior, the distribution deforms and so must its orthogonal polynomials. The quantum multinomial family is not a generalization of the classical multinomial in the sense of the Askey scheme (adding parameters or relaxing constraints); it is a parallel family, parametrized by the same quantities, which departs from classical statistics through the coherence of the amplitude summation. The classical-to-quantum distinction is thus orthogonal to the Askey hierarchy: the Askey scheme organizes distributions by parameter structure, while the quantum extension introduces a new axis, coherent versus incoherent combination, that the classical scheme does not capture. For $k = 2$, the answer is given in Sect.~\ref{sec:krawtchouk}: the quantum binomial distribution is the squared orthonormal Krawtchouk function, and its orthogonal polynomials are Christoffel transforms of the Krawtchouk family, computable via Darboux transformations. For $k \geq 3$, the amplitudes are genuinely complex and the squared-modulus operation breaks the linear structure; characterizing the resulting orthogonal polynomials remains open. A third direction concerns partial distinguishability. Real photons are never perfectly identical, and the transition from distinguishable to indistinguishable particles is governed by a partial distinguishability matrix~\cite{tichy2015,shchesnovich2015}. In the quantum multinomial framework, this transition interpolates between $\sum_{\Jb} w_{\Jb}\,|a_{\Jb}|^2$ (fully distinguishable, incoherent) and $|\sum_{\Jb} w_{\Jb}\, a_{\Jb}|^2$ (fully indistinguishable, coherent); the intermediate regime, where the coherent sum acquires off-diagonal damping factors from spectral overlaps, is a natural extension. The moment formulas of Sects.~\ref{sec:moments} and~\ref{sec:multiport-moments} also bear on the verification of boson sampling experiments: the $\binom{r}{j}^2$-versus-$\binom{r}{j}$ signature in the factorial moments reflects, at the level of efficiently computable low-order statistics, the same bra-ket pairing of amplitude expansions that gives rise to the permanent in the full output probability; together with the phase-dependent cross-mode covariance~\eqref{eq:cross-moment}, it provides low-order statistical witnesses that distinguish genuine multiphoton interference from classical or spoofed output distributions without requiring the full permanent computation~\cite{walschaers2016}. The third cumulant (Sect.~\ref{sec:multiport-cumulants}) is an additional such witness for multiport interferometers that is invisible in beam splitter calibration experiments, where $\kappa_3$ is invariant. More broadly, the quantum multinomial is defined in terms of standard combinatorial and linear-algebraic objects, without reference to Hilbert spaces or operator algebras; whether it arises in contexts beyond quantum optics remains to be explored.

\section*{Acknowledgements}
This work was supported by Grant PID2024-159557OB-C22 funded by MICIU/AEI/10.13039/501100011033/FEDER, EU. The authors acknowledge the use of Claude (Anthropic) for verification of mathematical derivations, computations, manuscript preparation, and proofreading.

\bibliographystyle{unsrt}
\bibliography{quantum_multinomial}

@article{reck1994,
  author  = {M. Reck and A. Zeilinger and H. J. Bernstein and P. Bertani},
  title   = {Experimental realization of any discrete unitary operator},
  journal = {Phys. Rev. Lett.},
  volume  = {73},
  pages   = {58--61},
  year    = {1994}
}

@article{scheel2008,
  author  = {S. Scheel},
  title   = {Permanents in linear optical networks},
  journal = {Acta Phys. Slovaca},
  volume  = {58},
  pages   = {675--810},
  year    = {2008}
}

@article{valiant1979,
  author  = {L. G. Valiant},
  title   = {The complexity of computing the permanent},
  journal = {Theor. Comput. Sci.},
  volume  = {8},
  pages   = {189--201},
  year    = {1979}
}

@article{aaronson2013,
  author  = {S. Aaronson and A. Arkhipov},
  title   = {The computational complexity of linear optics},
  journal = {Theory of Computing},
  volume  = {9},
  pages   = {143--252},
  year    = {2013}
}

@article{tichy2014,
  author  = {M. C. Tichy},
  title   = {Interference of identical particles from quantum to classical},
  journal = {J. Phys. B: At. Mol. Opt. Phys.},
  volume  = {47},
  pages   = {103001},
  year    = {2014}
}

@article{campos1989,
  author  = {R. A. Campos and B. E. A. Saleh and M. C. Teich},
  title   = {Quantum-mechanical lossless beam splitter: {SU(2)} symmetry and photon statistics},
  journal = {Phys. Rev. A},
  volume  = {40},
  pages   = {1371--1384},
  year    = {1989}
}

@book{mandel1995,
  author    = {L. Mandel and E. Wolf},
  title     = {Optical Coherence and Quantum Optics},
  publisher = {Cambridge University Press},
  year      = {1995}
}

@article{yurke1986,
  author  = {B. Yurke and S. L. McCall and J. R. Klauder},
  title   = {{SU(2)} and {SU(1,1)} interferometers},
  journal = {Phys. Rev. A},
  volume  = {33},
  pages   = {4033--4054},
  year    = {1986}
}

@article{hong1987,
  author  = {C. K. Hong and Z. Y. Ou and L. Mandel},
  title   = {Measurement of subpicosecond time intervals between two photons by interference},
  journal = {Phys. Rev. Lett.},
  volume  = {59},
  pages   = {2044--2046},
  year    = {1987}
}

@book{zare1988,
  author    = {R. N. Zare},
  title     = {Angular Momentum: Understanding Spatial Aspects in Chemistry and Physics},
  publisher = {Wiley},
  year      = {1988}
}

@article{koornwinder1982,
  author  = {T. H. Koornwinder},
  title   = {Krawtchouk polynomials, a unification of two different group theoretic interpretations},
  journal = {SIAM J. Math. Anal.},
  volume  = {13},
  pages   = {1011--1023},
  year    = {1982}
}

@incollection{deloera2014,
  author    = {J. A. {De Loera} and E. D. Kim},
  title     = {Combinatorics and geometry of transportation polytopes: an update},
  booktitle = {Discrete Geometry and Algebraic Combinatorics},
  series    = {Contemp. Math.},
  volume    = {625},
  publisher = {AMS},
  pages     = {37--76},
  year      = {2014}
}

@book{stanley2012,
  author    = {R. P. Stanley},
  title     = {Enumerative Combinatorics},
  volume    = {1},
  edition   = {2nd},
  publisher = {Cambridge University Press},
  year      = {2012}
}

@article{tichy2010,
  author  = {M. C. Tichy and M. Tiersch and F. {de Melo} and F. Mintert and A. Buchleitner},
  title   = {Zero-transmission law for multiport beam splitters},
  journal = {Phys. Rev. Lett.},
  volume  = {104},
  pages   = {220405},
  year    = {2010}
}

@article{tichy2015,
  author  = {M. C. Tichy},
  title   = {Sampling of partially distinguishable bosons and the relation to the multidimensional permanent},
  journal = {Phys. Rev. A},
  volume  = {91},
  pages   = {022316},
  year    = {2015}
}

@article{shchesnovich2015,
  author  = {V. S. Shchesnovich},
  title   = {Partial indistinguishability theory for multiphoton experiments in multiport devices},
  journal = {Phys. Rev. A},
  volume  = {91},
  pages   = {013844},
  year    = {2015}
}

@article{walschaers2016,
  author  = {M. Walschaers and J. Kuipers and J.-D. Urbina and K. Mayer and M. C. Tichy and K. Richter and A. Buchleitner},
  title   = {Statistical benchmark for {BosonSampling}},
  journal = {New J. Phys.},
  volume  = {18},
  pages   = {032001},
  year    = {2016}
}

@article{cardin2024,
  author  = {Y. Cardin and N. Quesada},
  title   = {Photon-number moments and cumulants of {Gaussian} states},
  journal = {Quantum},
  volume  = {8},
  pages   = {1521},
  year    = {2024}
}

@article{stobinska2019,
  author  = {M. Stobi\'nska and A. Buraczewski and M. Moore and W. R. Clements and J. J. Renema and S. W. Nam and T. Gerrits and A. Lita and W. S. Kolthammer and A. Eckstein and I. A. Walmsley},
  title   = {Quantum interference enables constant-time quantum information processing},
  journal = {Sci. Adv.},
  volume  = {5},
  pages   = {eaau9674},
  year    = {2019}
}

\end{document}